\documentclass[
showpacs,preprintnumbers,amsmath,amssymb]{revtex4}
\usepackage{amsmath}
\usepackage{float}
\usepackage{graphicx}
\usepackage[]{caption}
\usepackage{tensor}
\usepackage[toc,page]{appendix}
\usepackage{color,soul}

\DeclareMathOperator{\arctanh}{arctanh}

\begin{document}
\title{CHARACTERISTIC PROPERTIES OF TWO DIFFERENT VISCOUS COSMOLOGY MODELS FOR THE FUTURE UNIVERSE}

\author{Ben David Normann$^{1}$\footnote{Present affiliation: University of Stavanger, N-4036 Stavanger, Norway. Email: ben.d.normann@uis.no}  and Iver Brevik$^{2}$\footnote{E-mail:iver.h.brevik@ntnu.no}}

\medskip

\affiliation{$^{1}$Department of Physics, Norwegian University
of Science and Technology, N-7491 Trondheim, Norway}
\affiliation{$^{2}$Department of Energy and Process Engineering, Norwegian University
of Science and Technology, N-7491 Trondheim, Norway}

 \today

\begin{abstract}
We analyze characteristic properties of two different cosmological models: (i) a one-component dark energy model where the bulk viscosity $\zeta$ is associated with the fluid as a whole, and (ii) a two-component model where $\zeta$ is associated with  a dark matter component $\rho_{\rm m}$ only, the dark energy component considered inviscid. Shear viscosity is omitted. We assume throughout the simple equation of state $p=w\rho$, with $w$ a constant. In the one-component model we consider two possibilities, either to take $\zeta$ proportional to the scalar expansion (equivalent to the Hubble parameter), in which case the evolution becomes critically dependent on the value of the small constant $\alpha=1+w$ and the magnitude of $\zeta$. Second, we consider the case $\zeta=~$const., where a de Sitter final stage is reached in the future. In the two-component model we consider only the case where the dark matter viscosity $\zeta_{\rm m}$ is proportional to the square of $\rho_{\rm m}$, where again a de Sitter form is found in the future. In this latter case the formalism  is supplemented by a phase space analysis. As a general result of our considerations we suggest that a value $\zeta_0\sim 10^6~$Pa s for the present viscosity is reasonable, and that the two-component model seems to be favored.

\end{abstract}
 \maketitle

\section{Introduction}
\label{ChaptIntro}

Recent years have witnessed an increased interest in viscous cosmology. This is quite understandable -- at least so from a hydrodynamicist's point of view -- since the concept of an ideal (inviscid) fluid is after all an approximation to the real world, applicable under certain circumstances but in many cases definitely not. Some articles and reviews on viscous cosmology can be found in Refs.~\cite{weinberg71,gron90,zimdahl96,cardone06,mostafapoor11,brevikgron13,bamba15,brevik11,brevik12,brevik15,brevik15a,sasidharan15,normann16}. It is to be observed that a viscous fluid can also be understood as a class of inhomogeneous fluids, as introduced by Nojiri and Odintsov \cite{nojiri05}.

We will in this paper study two different models for the development of the viscous cosmic fluid in the late universe, i.e., the period from present time $t=0$ onwards. The late universe has of course been studied from various perspectives also before, then mostly within the context of ideal fluids. Especially the late universe shows rather dramatic properties if one takes it to be a one-component dark energy fluid. If $\rho$ denotes the dark energy density and $p$ the dark pressure one  knows that $p$ is negative. Moreover, if the equation-of-state parameter $w=p/\rho$ is less that $-1$ (a phantom fluid), then it has been known from the work of Caldwell et al. \cite{caldwell03} that the universe develops into a singularity after a finite time, called the big rip. Cf. also Refs.~\cite{nojiri03,nojiri04} (for classification of future singularities, see Ref.~\cite{nojiri05a}.
There exists other forms of the future singularity, especially the so-called little rip, for which the universe needs an infinite time to reach the singular state \cite{frampton11,brevik11,brevik12, frampton12a}.
Other variants are the pseudo-rip \cite{frampton12} and the quasi-rip \cite{wei12}.

Our intention to take  viscosity into account in the description of the future universe is to some extent motivated from the desirability to make the theory of the cosmic fluid more complete, and more importantly,  from  the amount of recent astronomical data. In order to give a reliable prediction about the future universe, one needs naturally accurate information about the values of present-day physical quantities, for our purpose especially the present bulk viscosity $\zeta_0$. There exists by now several works from which it is possible to at least pin down the interval of $\zeta_0$ to a reasonable range. In accordance with common usage we shall ignore the shear viscosity, this being motivated by the assumption of spatial isotropy on a large scale. We will in the following analyze the characteristics of two different models: first, a one-component dark energy model for which the viscosity is associated with the fluid as a whole;  second, a two-component model for which the viscosity is associated with the dark matter component only, the dark energy component remaining inviscid.

Before embarking on the formalism, we have to consider the equation of state. In this paper we will assume the simple homogenous equation
\begin{equation}
p=w\rho, \quad w=\rm{constant}
\label{1}
\end{equation}
throughout. There exists in the literature more complicated versions of this equation, but the simple form (\ref{1}) has the advantage that it makes the discussion of the influence from viscosity more transparent.

\section{One-component dark energy model}
Since the dark fluid is so dominant at present, and since observations suggest that this seems to pertain -- even in an increasing fashion -- a one-component model is arguably an appropriate model of the future universe. In this section we will construct such a model, with viscosity included. We assume a spatially flat FRW space, with metric
\begin{equation}
ds^2=-dt^2+a^2(t)d{\bf x}^2, \label{2}
\end{equation}
and put the cosmological parameter $\Lambda$ equal to zero. The quantities $\rho$ and $p$ refer to the full (dark) fluid.  With $\theta=3H=\dot{a}/a$ denoting the scalar expansion, we can write the Friedmann equations in the form
\begin{equation}
\theta^2=24\pi G\rho, \label{3}
\end{equation}
\begin{equation}
\dot{\theta}+\frac{1}{2}\theta^2=-12\pi G[p-\zeta(\rho)\theta], \label{4}
\end{equation}
and the energy conservation equation as
\begin{equation}
\dot{\rho}+(\rho+p)\theta =\zeta(\rho)\theta^2. \label{5}
\end{equation}
With the  constant $\alpha$ defined via
\begin{equation}
w=-1+\alpha, \label{6}
\end{equation}
we obtain the governing equation as
\begin{equation}
\dot{\rho}+\sqrt{24\pi G}\,\alpha \rho^{3/2}-24\pi G\zeta(\rho)\rho=0. \label{7}
\end{equation}
The solution is
\begin{equation}
t=\frac{1}{\sqrt{ 24\pi G}}\int_{\rho_0}^\rho \frac{d\rho}{\rho^{3/2}\left[ -\alpha +\sqrt{24\pi G}\,\zeta(\rho)/{\sqrt \rho}\right]}, \label{8}
\end{equation}
where  subscript zero refers to the present time $t=0$.

We now turn to the dependence of the bulk viscosity on the energy density. We will assume the form
\begin{equation}
\zeta(\theta)=\zeta_0\left( \frac{\theta}{\theta_0}\right)^{2\lambda}, \quad \lambda = \rm{constant}. \label{9}
\end{equation}
This choice might seem more arbitrary than it is. Using Friedmann's first equation we rewrite it as
\begin{equation}
\zeta(\rho)=\zeta_0\left(\frac{\rho}{\rho_0}\right)^{\lambda} \label{10}
\end{equation}
{which might seem more natural. Indeed, such a dependency of $\zeta$ on the energy density to some power $\lambda$ has been investigated in many earlier works, cf. e.g.} Refs.~\citep{murphy73, barrow86, nojiri05a, gorbunova05, brevik06, campo07, li10, paolis10, velten12, velten13, Frontiers, cardenas15, normann16}.

 Let us  consider the ansatz \eqref{10} in some more detail. Dealing with the dark sector, it is inevitable that some guesses have to be made since so little is known. Aiming at a reasonable simple formalism, and assuming that viscosity has an underlying microscopic basis, it is natural to assume that the viscosity either is a  constant, or  it depends on one of  the thermodynamic variables of the system.  Whether we take this variable to be the temperature $T$ or the energy density $\rho$ seems to be of less importance, since these quantities are interrelated. The choice of temperature might be natural for the early universe, where the temperature dependence upon  redshift is known for the different fluid components (that is to say, the value of $\alpha_{\rm i}$ is known in the formula $T\,\sim\,(1+z)^{\alpha_{\rm i}}$ for  different components $i$). Several previous works have taken this approach (e.g. ~\citep{zimdahl96, gagnon11, wang14}. As the temperature concept becomes a delicate issue in the dark sector,  it appears  that the previous literature has made use of  Equation \eqref{10} with different choices for $\lambda$ in an attempt to minimize the arbitrariness of $\zeta$. Also note that Eq.~\eqref{10} nevertheless incorporates a wide spectrum of scenarios (even the constant case; i.e. $\lambda\,=\,0$). Note for instance the (dynamically) close resemblance with the Chaplygin gas (for $\lambda=-1$, $w=0$) and modified Chaplygin gas (for $-1\,\leq\,\lambda\,\leq 0$). Refer, for instance, to Refs. \citep{pourhassan13, saadat13, naji14, naji14b, bennaoum14} for works relating to the Chaplygin gas. In all, powers of $\lambda$ ranging in the interval $-1\leq \,\lambda\,\leq\,1$ have been widely used in the literature, either formulated as a modification of the equation of state, or by introducing viscosity.

In our previous works,  we have  investigated the choices $\lambda=0,\,1/2,\,$and $1$.\cite{normann16} All three choices seem to suggest the same order of magnitude for $\zeta_0$. In the following we shall concentrate on $\lambda=0$, justified by its simplicity, and on $\lambda=1/2$. The last choice is made because this viscosity can always  be interpreted as a linear modification of the equation-of-state parameter, such as $w\,\rightarrow\,w'$, where both $w$ and $w'$ are constants; cf. Ref.~\cite{normann16} for details. We could have employed more exotic forms of the viscosity, as is sometimes done  (refer to the review of Dou and Weng \cite{dou11}).  However, we will not introduce such complicated assumptions in this paper.

  We shall now leave the qualitative aspects of the choice of viscosity, and instead consider  quantitative parts of our preferred alternatives.

\bigskip

 {\it $(i) \lambda=1 ~(\zeta \propto \sqrt{\rho}$).} In this case Equation (\ref{8}) yields the solution
\begin{equation}
t=\frac{1}{\theta_0}\,\frac{2}{X_0}\left(1-\frac{1}{\sqrt{ \Omega}}\right). \label{11}
\end{equation}
where $\Omega=\rho/\rho_{\rm c}$ denotes relative density. Here $\rho_{\rm c}=\rho_0$ since $k=0$. Also, we have for convenience introduced the notation
\begin{equation}
 X_0=\frac{2\,B}{\theta_0}-\alpha\label{X0}
\end{equation}
with the quantity $B$ defined as
\begin{equation}
B=12\pi G \zeta_0. \label{16a}
\end{equation}

The remarkable property of the expression (\ref{11}), as discussed also earlier \cite{gorbunova05,brevik15, normann16}, is that it permits a big rip singularity to occur even if the fluid is initially in the quintessence region $\alpha >0$. The condition for such a singularity is simply
\begin{equation}
X_0>0.
\end{equation}
If this condition holds, the  singularity ($\rho=\infty$) occurs at a finite  time
\begin{equation}
t_{\rm s}= \frac{1}{\theta_0}\,\frac{2}{X_0}, \quad (\zeta \propto \sqrt{\rho}). \label{12}
\end{equation}
By contrast, if  $X_0<0$,  the universe becomes infinitely diluted, $\rho \propto 1/t^2$, when $t\rightarrow \infty$.

\bigskip
{\it $(ii) \lambda=0~(\zeta={\rm constant}=\zeta_0$).} From Eq.~(\ref{8}) we then have
\begin{equation}
t=\frac{1}{\sqrt{ 24\pi G}}\int_{\rho_0}^\rho \frac{{\rm d}\rho}{\rho^{3/2}\left[ -\alpha +\zeta_0 \sqrt{24\pi G/\rho}\right]}. \label{13}
\end{equation}
Consider first as a separate case an inviscid fluid, $\zeta_0=0$,
\begin{equation}
t=-\frac{1}{\sqrt{24\pi G}}\frac{1}{\alpha}\left( \frac{1}{\sqrt{\rho_0}}-\frac{1}{\sqrt{\rho}}\right)=-\frac{1}{\theta_0\alpha}\left(1-\frac{1}{\sqrt{\Omega}}\right)\label{14}
\end{equation}
Thus if $\alpha<0$ (phantom region), one gets $\rho \rightarrow \infty$ after a finite time
\begin{equation}
t_{\rm s}=-\frac{1}{\theta_0\alpha}, \quad (\zeta_0=0).  \label{15}
\end{equation}
If $\alpha >0$, one gets $\rho \rightarrow 0$ as $t\rightarrow \infty$. The behaviour in this case has been  known since the discovery of Caldwell et al. \cite{caldwell03}.

Assume  next that the viscosity $\zeta_0$ is nonzero and finite. We then get as solution
\begin{equation}
t=\frac{1}{B}\ln \left[ \frac{X_0}{-\alpha + 2B/(\theta_0\sqrt{\Omega})}\right], \quad (\zeta=\zeta_0). \label{16}
\end{equation}
We then obtain for the energy density
\begin{equation}
\Omega=\frac{\rho}{\rho_0}=\left[ \frac{\alpha +X_0}{\alpha+X_0e^{-Bt}}\right]^2. \label{16b}
\end{equation}
This means that the solution is of the de Sitter type: in the far future, $\rho \rightarrow~$const, which implies $\theta \rightarrow~$const. Denoting the limiting value of $\rho$ by $\rho_{\rm dS1}$ (the extra subscript 1 referring to model 1), we get
\begin{equation}
\rho_{\rm dS1}=\rho_0\left( 1+\frac{X_0}{\alpha}\right)^2=\frac{24\pi G\zeta_0^2}{\alpha^2}. \label{16c}
\end{equation}
It is thus apparent that the future fate of the universe will in general depend on {\it both} parameters $\alpha$ and $X_0$:

\bigskip

\noindent 1. If both $\alpha>0$ and $X_0>0$, then $\rho_{\rm dS1}>\rho_0$.

\noindent 2. If $\alpha>0$ and $X_0<0$, then  $\rho_{\rm dS1}<\rho_0$.

\noindent 3. If $\alpha <0$ then $X_0 >0$ always, and $\rho_{\rm dS1}<\rho_0$.

\bigskip

In all cases, the limiting value is given by the same expression (\ref{16c}).  (This might  appear surprising, but one notices that for a given value of $\rho_0$,   $X_0>0$ and $X_0<0$  correspond to different values of $\alpha$.)

 The situation is reminiscent of  the {\it little rip} case, the difference being that the limiting value for the density obtained after an infinite span of time is  finite value  instead of an infinite one. This has previously been defined as a pseudo-rip.~\cite{frampton12}.

 Although we postpone the main part of the numerical discussion to Section IV, it is convenient here to bring in some numbers from the observations. From the 2015 Planck data, Ref.~\cite{ade15} Table 5, we have $w=-1.019^{+0.075}_{-0.080}$. Thus, $\alpha =1+w$ will be lying within two limits,
 \begin{equation}
 \alpha_{\rm min}=-0.099, \quad \alpha_{\rm max}=+0.056. \label{16d}
 \end{equation}
 Taking $\zeta_0=10^6~$Pa s as a reasonable mean value for the present viscosity, we then evaluate the quantity $B=12\pi G \zeta_0$ to be about 1 km s$^{-1}$Mpc$^{-1}$ in astronomical units. With $H_0=67.74$  km s$^{-1}$Mpc$^{-1}$ we get $2B/\theta_0=0.00984$ as an estimate. Then, according to Eq.~(\ref{X0}) we have
 \begin{equation}
 X_0(B=1, \alpha_{\rm max})=0.00984-\alpha_{\rm max}=-0.0462,
 \end{equation}
 \begin{equation}
 X_0(B=1, \alpha_{\rm min})=0.00984-\alpha_{\rm min}=+0.109.
 \end{equation}
This means that we recover the cases 2 and 3 above; the future de Sitter energy density will become lower than the present one.

 Before leaving this section we ought to make a remark regarding the physical meaning of the viscosity $\zeta_0$ when it is assumed to be a constant. As we will discuss below, the magnitudes of $\zeta_0$ inferred from experiments are many orders of magnitude greater than the viscosities in ordinary hydrodynamics. Under such circumstances, especially under extreme conditions such as close to the big bang, one should expect that the viscosity refers to a turbulent state of motion. Now, in ordinary hydrodynamics the usual recipe for constructing the turbulent viscosity is to assume it proportional to the Prandtl mixing length.  Here there is only one length parameter, namely the scale factor $a$. The question is: does this circumstance come into conflict with the constant viscosity assumption? In our opinion the answer is no. The cosmology problem is fundamentally different from the ordinary kinetic theory one because the geometry itself is changing. Moreover, the values of the parameter $w$ in cosmology often do not fit the usual fluid dynamical conditions. We thus conclude that the assumption about a constant viscosity in the cosmology case, even under turbulent conditions, is not physically excluded after all.

\section{Two-component model}

Our second model is motivated by the fact that very little is known about the physics of the dark fluid. It is questionable if we can  associate a bulk viscosity with it at all. We may here refer to two previous works; Refs.~\cite{velten12} and \cite{gagnon11}, The former  is phenomenological in nature, whereas the latter presents a microscopic model of dark energy and bulk viscosity. We find it most simple to keep  the discussion on a phenomenological level. Comparison  with observations means that we  look back in time, when matter was increasingly more dominant. In view of the uncertainty, a two-component model appears to be physically  suitable. We suggest the following  alternative picture for the cosmic fluid:

\noindent (i)  Assume that the fluid consists of two components, one dark matter component with density $\rho_{\rm m}$, and one dark energy component described entirely through the cosmological constant $\Lambda$.

\noindent (ii) Assume that the dark matter is viscous, described by the same form for $\zeta$  as above, and let the dark energy component be wholly ideal.

\bigskip
Note that this model reduces to the $\Lambda$CDM model for $\zeta=0$.
Let us look at the formal consequences of these assumptions. The kind of Friedmann equations that we have to do with now (still with curvature parameter $k=0$), are
\begin{equation}
\theta^2=24\pi G\left( \rho_{\rm m}+\frac{\Lambda}{8\pi G}\right), \label{18}
\end{equation}
\begin{equation}
\dot{\theta}+\frac{1}{2}\theta^2=-12\pi G\left( p_{\rm m}-\zeta_{\rm m} \theta -\frac{\Lambda}{8\pi G}\right), \label{19}
\end{equation}
and the energy conservation equation for the matter part is
\begin{equation}
\dot{\rho}_{\rm m}+(\rho_{\rm m}+p_{\rm m})\theta =\zeta_{\rm m} \theta^2. \label{20}
\end{equation}
We will model the viscosity of dark matter with the same form as Eq.~(\ref{10}), though we must replace $\rho$ with $\rho_{\rm m}$. $\zeta_0$ is as before taken to be the present-time measured bulk viscosity.  (Note that the previous alternative form in Eq.~(\ref{9}) is not any longer an equivalent option, due to the generalized Friedmann's equation (\ref{18}).)

 As for the equation of state we will assume, in analogy with Eq.~(\ref{1}), that
 \begin{equation}p_{\rm m}=w_{\rm m}\rho_{\rm m}, \quad w_{\rm m}={\rm constant}. \label{23}
 \end{equation}
 and as in the previous case, we assume that the viscosity option of main interest is when $\lambda=1/2$, i.e.
  \begin{equation}
  \label{darkvisc}
  \zeta_{\rm m}=\zeta_0\left(\frac{\rho_{\rm m}}{\rho_{\rm m 0}}\right)^{1/2}=\zeta_0\left(\frac{\theta^2-3\Lambda}{\theta_0^2-3\Lambda}\right)^{1/2}.
  \end{equation}
We will here restrict ourselves to this option. We obtain from this the  governing equation for the scalar expansion

 \begin{equation}
\dot{\theta}=\frac{1+w_{\rm m}}{2}\left[ 3\Lambda-\theta^2+\frac{24\pi G\zeta_0}{1+w_{\rm m}}\left(\frac{\theta^2-3\Lambda}{\theta_0^2-3\Lambda}\right)^{1/2}\theta\right], \label{24}
 \end{equation}
 which may be rewritten as
 \begin{equation}
\label{GovEq}
\dot{\theta}=\Gamma\, \theta\, \sqrt{\theta^2 - 3\Lambda}-\phi\, \left(\theta^2-3\Lambda\right),
\end{equation}
through the definitions
\begin{equation}
\label{phiAndGamma}
\Gamma\equiv\frac{B}{\sqrt{\theta_0^2-3\Lambda}}=\frac{B}{\sqrt{\Omega_{\rm 0m}}\theta_0} \phantom{000}{\rm and}\phantom{000}\phi=\frac{1+w_{\rm m}}{2},
\end{equation}
where $\Omega_{\rm 0m}=1-\Omega_\Lambda, \,
\Omega_\Lambda=3\Lambda/\theta_0^2$.

The integral solution for $t$ is
 \begin{equation}
 t=\int_{\theta_0}^{\theta}{\frac{1}{\Gamma\, \theta\, \sqrt{\theta^2 - 3\Lambda}-\phi\, \left(\theta^2-3\Lambda\right)}}\rm d\theta
 \label{25}
 \end{equation}
 in the future universe. Solving the integral one finds
\begin{equation}
\label{partSolt2}
t(\theta,\phi,\Gamma)=\frac{1}{\theta_0\sqrt{\Omega_\Lambda}\sqrt{\phi^2-\Gamma^2}}\left[G(\theta,\phi,\Gamma)-G(\theta_0,\phi,\Gamma)\right]
\end{equation}
where the function $G$ was introduced for convenience of notation only. It is defined such that
\begin{equation}
\label{A}
G(\theta,\phi,\Gamma)\equiv\arctanh\!\left(\sqrt{\frac{\theta^2}{\theta_0^2\Omega_\Lambda}\left( 1-\frac{\Gamma^2}{\phi^2}\right)}\right) + \arctanh\!\left(\sqrt{\left(\frac{\theta^2}{\theta_0^2\Omega_\Lambda}-1\right)\left(\frac{\phi^2}{\Gamma^2}-1\right)}\right).
\end{equation}
Eq.~\eqref{partSolt2} is a solution provided that $\Gamma^2-\phi^2\theta^2+3\Lambda\phi^2\,\neq\,0$, or, equivalently; $(\Gamma /\phi )^2\,\neq\,\Omega_{\rm m}\theta_0^2$. Since $\Gamma\, < \, \phi$ for the regimes we are looking at, and since $\theta_0^2\,\Omega_{\rm m}\,>\, 1$ for the same regimes, this condition is fulfilled.

\subsection*{Phase space analysis of the two-component model}

\begin{figure}[H]
\centering
\includegraphics[width=0.8\linewidth]{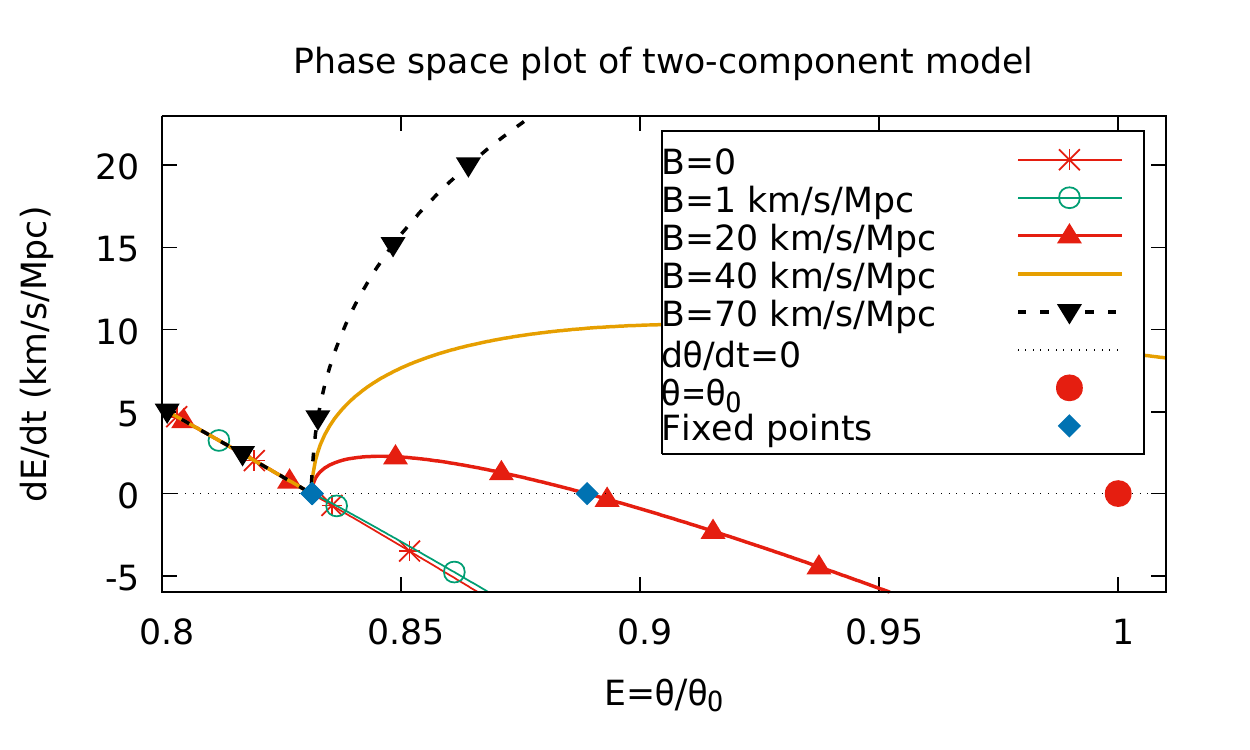}
\caption{In the plot $E=\theta/\theta_0=H/H_0$ is the dimensionless Hubble parameter. Different colours show $\dot{E}(E)$ for increasing viscosities in the range $B\,=0$ to $70\,\rm\,km\,s^{-1}\,Mpc^{-1}$. The red filled circle to the right marks the position of the present universe; $E_0=1$. Black, dotted horizontal line shows $\dot{\theta}=0$. Fixed points are marked with blue, tilted squares. The left-most blue, tilted square shows the fixed point $\theta_{\rm dS2}=\theta_0\sqrt\Omega_\Lambda$. Note that since $dE/dt\,<0$ at present, the universe is pushed towards a stable attractor at $\theta_{\rm dS2v}$. As can be seen from the plot, the function of viscosity is to increase the value of $\theta_{\rm dS2v}$; i.e.; to enhance the development towards a de Sitter state. The fixed point $\theta_{\rm dS2}$ corresponding to $B\,=\,1\,\rm\,km\,s^{-1}\,Mpc^{-1}$ is indistinguishable from the fixed point at $\theta=\theta_0\sqrt\Omega_\Lambda$.}
\label{phasespace}
\end{figure}

\noindent At this point it is instructive to carry out a phase space analysis to locate potential fixed points for $\theta$. A fixed point is characterized by $\dot{\theta}=0$ and is   stable if $\rm d\dot{\theta}/d\theta\,< 0$.  Denoting fixed points by $\theta_{\rm fp}$, we have the conditions
\begin{equation}
\dot{\theta}\bigg|_{\theta_{\rm fp}}=0\phantom{000}\text{and}\phantom{000}\frac{\rm d\dot{\theta}}{\rm d\theta}\bigg|_{\theta_{\rm fp}}\,<0\,\phantom{000}\Longleftrightarrow\phantom{000}\theta_{\rm fp}\,\text{ is a  de Sitter solution.}
\end{equation}
In other words, a stable fixed point represents a de Sitter evolution.
Applying  the condition $\dot{\theta}=0$ on Eq.~\eqref{GovEq} we find two sets of fixed points, given by
\begin{equation}
\label{fp1}
\theta_{\rm dS2}^2=3\Lambda=\Omega_\Lambda\theta_0^2\phantom{000}\text{(inviscid f.p.)}\phantom{000}\text{or}\phantom{000}\theta_{\rm dS2v}^2=    \frac{\Omega_{\Lambda}\theta_0^2}{1-\left(\Gamma/\phi\right)^2}\phantom{000}\text{(viscous f.p.)},
\end{equation}
where $\theta_{\rm dS2}$ and $\theta_{\rm dS2v}$ are fixed points. Subscript 'dS'  indicates that a universe approaching one of these fixed points approaches a de Sitter state, while subscript '2' is to distinguish from the one-component models in the previous section. The rightmost solution in the above equation, $\theta_{\rm dS2v}$, bears the subscript 'v' to specify that  these fixed points are viscosity-dependent. Calculating d$\dot{\theta}/\rm d\theta$ at the fixed points one finds that $\theta_{\rm dS2}$ is stable if $B=0$ and that $\theta_{\rm dS2v}$ is stable for all $\Gamma\,<\,\phi$. Thus the position of the fixed point  $\theta_{\rm dS2}$ is independent of the viscosity, but its stability is viscosity dependent. It will be stable  only if there is no viscosity; see Fig.~\ref{phasespace} for details. The fixed points to the right are de Sitter universes since they are stable. These  exist only for $\Gamma\,<\,\phi$,  since this will leave $\theta_{\rm dS2}$ real and finite (we restrict ourselves to $\Gamma\,>\,0$). The condition $\Gamma\,<\,\phi$ corresponds to $B<56.47\,\rm km\,s^{-1}Mpc^{-1}$.

\paragraph*{}The two-component model studied in the present section is of most interest for predictions about the future universe. As it stands, however, it  clearly resembles the $\Lambda \rm CDM$ model and should accordingly be expected to be valid back in time, deep into the matter dominated epoch since we have included a matter and a constant dark energy term. Furthermore, since the viscosity-affected fixed points (to the right in Eq.~\eqref{fp1}) are stable, we should require that {\it  no fixed point occurs before present time}. To understand this let us assume the opposite: if $t=t_0$ denotes the present time, imagine that  there is a stable fixed point located somewhere at $t<t_0$. This  means that the universe reaches a de Sitter stage at some time $t'$ in the past. Since the universe cannot pass through such a stable fixed point, it will  have  to remain there. That is to say,  $\dot{\theta}(t')=0$. This appears incompatible with observations which show that $\theta (t)\,\neq\,$ const. for $t\,\leq t_0$. Since the scalar expansion was greater in the past (i.e. $\theta(t<t_0)\,>\,\theta_0$), the requirement that no stable fixed point is allowed to occur in the past may be written as
\begin{equation}
\label{fp2}
\frac{\theta_{\rm dS2v}^2}{\theta_0^2}=\frac{\Omega_{\Lambda}}{1-\left(\Gamma /\phi\right)^2}<\,1.
\end{equation}
This can be reformulated as
\begin{equation}
\label{req1}
\Gamma\,<\phi\,\sqrt{\Omega_{\rm 0m}}\,\phantom{000}\rightarrow\,\phantom{000}B<31.39\,\rm\,km\,s^{-1}\,Mpc^{-1}\,\phantom{000}\rightarrow\,\phantom{000}\zeta_0 <3.6\cdot 10^7\rm Pa~s.
\end{equation}
At this point we take the opportunity to mention that the reason why no such requirement was made in the one-component model in the last section, was that the one-component model is meant to be a good description of the future universe, and not the past.

From $\theta^2=\theta_0^2(\Omega_{\rm m}+\Omega_\Lambda)$ it follows from Eq.~\eqref{fp1} that the de Sitter value of the matter component becomes
\begin{equation}
\label{deSmatter}
\frac{\rho_{\rm m\, dS}}{\rho_{0\rm m}}=\frac{\Omega_{\rm m\, dS}}{\Omega_{0\rm m}}=\frac{\Omega_\Lambda}{\Omega_{0\rm m}}\frac{1}{\left(\phi\,/\,\Gamma\right)^2-1}
\end{equation}
The de Sitter value of the scalar expansion is correspondingly found directly from Equation~\eqref{fp1} for $\theta_{\rm dS2v}$. Note that in the case of $B\,\rightarrow\,0$, $\Omega_{\rm m\, dSv}\,=\,\Omega_{\rm m\, dS}\,\rightarrow\,0$.
We turn to numerical considerations in the next section.
\paragraph*{} In all, therefore, the fixed point analysis reveals an upper bound on the viscosity, by the requirement that no fixed point has occurred in the matter and dark energy dominated past. As we shall see in the next section, the numerical estimates of $B$ lies below this region. In concluding this section we  add a few  remarks:
\paragraph*{}(i) In general, only \textit{one} true de Sitter expansion can be explained by our present one-dimensional phase space model. This must be so because the universe cannot pass through a stable fixed point.
\paragraph*{}(ii) The present model does not have the capacity to account for an early inflationary phase. To do so in a one-dimensional model, an initial unstable fixed point would be required, where $\dot{\theta}$ transitions from positive to negative (or a bottle neck). As shown by Eq.~\eqref{fp1}, however, no such fixed point or bottle necks exist in our model. This does not -- at least not directly -- affect the present purpose, which is the universe from the matter dominated epoch and onwards (and especially the future universe), for which our model is expected to become progressively more correct.
\paragraph*{}(iii) Note that the natural parameter in this theory is $\phi\,/\,\Gamma$. As  pointed out several times in previous investigations, and most lately in~\cite{normann16}, this represents a degeneracy in the model. One is confronted with a choice:  either modify the equation-of-state parameter, or  include viscosity. Our choice has been to take  $w=0$ for matter in accordance with the standard model, and to introduce  viscosity.

\section{Numerical considerations }
\label{Sec:NumCons}
 It is important to know the approximate magnitude of the present day viscosity $\zeta_0$. There exist several recent investigations of this problem, most of them suggesting that $\zeta_0$ lies in the interval from $10^5$ to about $10^7$ Pa s. These values are of course many orders of magnitude higher than those encountered under daily circumstances. The analysis of Wang and Meng is useful \cite{wang14}, as it compares the theoretical curve for $H(z)$ as a function of the redshift $z$ with a number of observations. As a (logarithmic) mean of their observations we infer $\zeta_0 \sim 10^5$ Pa s. The analysis of Sasidharan and Mathew \cite{sasidharan15} leads to a value for $\zeta_0$ that is somewhat larger, about $7\times 10^7$ Pa s. This is roughly in agreement also with Velten and Schwarz \cite{velten12}.  Recent  considerations by us  led to comparable results \cite{brevik15,normann16, brevik16}. However, these values are seen to be somewhat large in comparison to our predictions above (cf. Eq.~\eqref{req1}). As mentioned in Section II we make use of the 2015 Planck data \cite{ade15}, from which we find that  $\Omega_\Lambda=0.6911$ and $\Omega_{\rm DM}=0.3089$.

\paragraph*{}Now go back to our  previous work \cite{normann16}, in which  we found solutions of the energy conservation equation for bulk viscous Friedmann universes with $k=0$. From that we conclude that a reasonable mean value for the current bulk viscosity is
\begin{equation}
\zeta_0 = 10^6~\rm Pa~s.
\end{equation}
This value is based upon calculations involving a multi-component fluid, but with an overall phenomenological viscosity. Since the dark fluid seems to become increasingly dominant in the future universe, both the multicomponent and one-component models will typically become more and more similar, as one component comes to dominate. This suggests that the above value is a reasonable estimate to go by -- at least for the one-component model. Using this value, we can give some quantitative predictions for the future universe. For convenience we will in this section use dimensional units. Thus the quantity $B$ is now written
\begin{equation}
B=12\pi G \zeta_0/c^2.
\end{equation}
As $B$  is often expressed in astronomical units, km s$^{-1}$Mpc$^{-1}$, we note the following useful formula relating $B$  to the viscosity in standard physical units,
\begin{equation}
\label{C_B_to_Zeta}
\zeta_0 = B{\rm [astro.units]}\times 1.15\times 10^6 \text{ Pa s}.
\end{equation}

We now start from $12\pi G/c^2=2.79\times 10^{-26}~$m/kg, and the value   $H_0=67.74$ km s$^{-1}$ Mpc$^{-1}= 2.20\times 10^{-18}$ s$^{-1}$ of the Hubble parameter at present. Identifying $\rho_0$ with the critical density $\rho_{\rm c}=2\times 10^{-26}$ kg m$^{-3}$ (setting the conventional $h$ parameter equal to 0.7), we obtain the following relative de Sitter energy density in model 1, Eq.~(\ref{16c}),
\begin{equation}
\frac{\rho_{\rm dS1}}{\rho_0}=\frac{24\pi G/c^2}{\rho_0} \left(\frac{\zeta_0}{\alpha c}\right)^2 \approx \frac{3}{\alpha^2}\times 10^{-5}.\label{forhold}
\end{equation}
The evolution of the universe is thus in this model critically dependent on the value of $\alpha$. If $\alpha < 0.0055$ the density increases into the future, whereas if $\alpha > 0.0055$ the density decreases.  If we raise $\zeta_0$ to $10^7~$Pa s, then the first restriction on $\alpha$ is relaxed, to $\alpha <0.05$ approximately. Note that while the last equation involves $|\alpha|$ only, the case $\alpha <0$ always leads to $\rho_{\rm dS1}> \rho_0$ as shown in Sect. II.

Consider now model 2. A viscosity $\zeta_0\,\sim\,10^6$ Pa s corresponds to $B\,\sim\,1\rm\,km\,s^{-1}\,Mpc^{-1}$. As mentioned, several other investigations suggest numbers in the same regime. As the models considered in the  literature are not identical, some  discrepancies in  numbers are expected. Generally, the more components introduced into the description (the more accurate description), the less  viscosity should be needed to account for the behaviour of the fluid. We therefore expect  that the viscosity in model 2 is at least no greater than $B=1\,\rm\,km\,s^{-1}\,Mpc^{-1}$. Inserting $B=1\,\rm\,km\,s^{-1}\,Mpc^{-1}$ into Eq.s~\eqref{fp1} and \eqref{deSmatter}  one finds
\begin{equation}
\label{deSmatter2}
\frac{\theta_{\rm dS2}^2}{\theta_0^2}\,\approx\,0.69\phantom{000}\text{and}\phantom{000}\frac{\rho_{\rm m\, dS2}}{\rho_{0\rm m}}\,\approx\,7.0\cdot 10^{-4},
\end{equation}
respectively. Working out the condition for $\Omega_{\rm m\, dS2v}\,<\Omega_{0\rm m}$ one  again finds $B\,<\,\sqrt{\Omega_{0\rm m}}\,\phi\,\theta_0=31.39\,\rm\,km\,s^{-1}\,Mpc^{-1}$. Reducing the viscosity to $\zeta_0\,\sim\,10^5$~Pa~s would similarly lower the matter ratio by one order of magnitude. This is still distinctively different from the inviscid case, where $\Omega_{\rm m\, dS2}=0$. The scalar expansion ratio is roughly unchanged. Correspondingly, $\zeta_0\,\sim\,10^7$ Pa s gives $\rho_{\rm m\, dS2v}\,/\,\rho_{0\rm m}\,\approx\,0.072$ and $\theta_{\rm dS2v}^2\,/\,\theta_0^2\,\approx\,0.71$. Finally for model 2, the characteristic time-scale is numerically found to be $\tau\,=\,5,68\cdot\,10^{-3}\rm\,s\,Mpc/km$, which gives
\begin{equation}
\label{tau}
\tau=5.56\rm\,Gy.
\end{equation}

\section{Conclusion}
Assuming a spatially flat FRW universe, we have been considering two viscous cosmology models, one one-component model with governing equations (\ref{3})-(\ref{4}), and one two-component model with governing equations (\ref{18})-(\ref{19}).   In all cases, we put the equation-of-state parameter $w$ equal to $-1+\alpha$, with $\alpha$ a positive or negative small constant.

 In  model 1 (Sect. II)   we considered two different options for the bulk viscosity $\zeta$. The first of these put the viscosity   proportional to the scalar expansion $\theta$,
\begin{equation}
\zeta =\zeta_0\theta/\theta_0,
\end{equation}
so that $\theta=\theta_0$ at present time $t=0$. In this case we found that a big rip singularity will appear in the far future if the parameter $X_0$ in Eq.~(\ref{X0}) is greater than zero. Both quantities $\alpha$ and $\zeta_0$ are thus of importance here. If $X_0<0$, the universe becomes infinitely diluted in the future.

The second option in model 1,
\begin{equation}
\zeta=\zeta_0=\rm constant,
\end{equation}
was found to give a de Sitter solution in the far future. Again, a combination of $\alpha$ and $X_0$ was found to be essential, for whether the density $\rho$ is to increase or decrease with time. For $\zeta_0\,\sim\,10^6\,\rm Pa~ s$ it would decrease in the case of a de Sitter future, but the possibility for a big rip remained, depending on the value of $\alpha$.

The two-component model (model 2) considered in Sect. III, corresponding to a viscosity associated with the matter component $\rho_{\rm m}$ only,
\begin{equation}
  \zeta_{\rm m}=\zeta_0\left(\frac{\rho_{\rm m}}{\rho_{\rm m 0}}\right)^{1/2}
  \end{equation}
was again found to inevitably lead to a de Sitter future, since the requirement that no stable fixed point is allowed to have occurred in the past, had to be made. I. e.; in contrast to the one-component model, the two-component model did not contain a big rip solution. The characteristic time-scale in this model was given by Eq.~\eqref{tau}. For the magnitude of viscosity permitted by the theory the scalar expansion $\theta$ decreases with time.

To help distinguishing between the various theoretical options, observational information was exploited in Sect. IV. Especially the value of $\zeta_0$ was of primary interest. Making use of information from various sources, we suggested that the value
\begin{equation}
\zeta_0=10^6~\rm Pa~s
\end{equation}
is a reasonable mean value. Using this, we found in Eq.~(\ref{forhold}) the de Sitter energy density of model 1 to be critically dependent on the value of $\alpha$. In model 2, the entity $\left(\phi/\Gamma\right)^2$ was of major importance, though not as critical as $\alpha$ in model 1. In model 2 we found the de Sitter scalar expansion to be about $\,30\%$ lower than $\theta_0$; cf. Eq.~(\ref{deSmatter2}).

Although all  cases covered in our discussions  may be compatible  with reality, we find that a two-component model (matter and dark energy) seems easiest to reconcile with the observations. A possible next step with regard to the one-component model might  be to extend it to a two-component model with an overall viscosity and examine its phase space. The equations for such a model were actually worked out in our previous paper ~(\cite{normann16}, Eqs. (31)), which could serve as a starting point for such an analysis. However, at present we restrict ourselves to expressing our preference for  the two-component model.



\end{document}